\documentclass[10pt,conference]{IEEEtran}
\IEEEoverridecommandlockouts
\usepackage{cite}
\usepackage{amsmath,amssymb,amsfonts}
\usepackage{algorithmic}
\usepackage{graphicx}
\usepackage{textcomp}
\usepackage{xcolor}
\usepackage{color}
\usepackage{tcolorbox}
\usepackage{booktabs}
\usepackage{balance}
\usepackage{tabularx}
\usepackage[hyphens]{url}

\newcommand{\mcode}[1]{$\tt #1$}

\def\BibTeX{{\rm B\kern-.05em{\sc i\kern-.025em b}\kern-.08em
    T\kern-.1667em\lower.7ex\hbox{E}\kern-.125emX}}
\begin{document}

\title{RAID: Tool Support for  Refactoring-Aware \\ Code Reviews}

\author{
\IEEEauthorblockN{Rodrigo Brito}
\IEEEauthorblockA{ASERG Group - Department of Computer Science \\
Federal University of Minas Gerais (UFMG)\\
Belo Horizonte, Brazil \\
britorodrigo@dcc.ufmg.br}
\and
\IEEEauthorblockN{Marco Tulio Valente}
\IEEEauthorblockA{ASERG Group - Department of Computer Science \\
Federal University of Minas Gerais (UFMG)\\
Belo Horizonte, Brazil \\
mtov@dcc.ufmg.br}
}

\maketitle

\begin{abstract}
Code review is a key development practice that contributes to improve software quality and to foster knowledge sharing among developers. However, code review usually takes time and demands detailed and time-consuming analysis of textual diffs. Particularly, detecting refactorings during code reviews is not a trivial task, since they are not explicitly represented in diffs. For example, a Move Function refactoring is represented by deleted (-) and added lines (+) of code which can be located in different and distant source code files. To tackle this problem, we introduce RAID, a refactoring-aware and intelligent diff tool. Besides proposing an architecture for RAID, we implemented a Chrome browser plug-in that supports our solution. Then, we conducted a field experiment with eight professional developers who used RAID for three months. We concluded that RAID can reduce the cognitive effort required for detecting and reviewing refactorings in textual diff. Besides documenting refactorings in diffs, RAID reduces the number of lines required for reviewing such operations. For example, the median number of lines to be reviewed decreases from 14.5 to 2 lines in the case of move refactorings and from 113 to 55 lines in the case of extractions.
\end{abstract}


\section{Introduction}

Code review is a widely used software engineering practice \cite{bacchelli2013, Bacchelli2018}. It has its origins in the 70s, when it was performed according to formal and strict guidelines, such as the ones proposed by Fagan {\em et al.} \cite{fagan1976}. Over the years, lightweight code review practices have emerged and gained popularity, in order to make the process more agile. Nowadays, software companies of all sizes and impact require their developers to engage in code reviews. For example, ``code review is one of the most important and critical processes at Google'' \cite{googleSE2020}. Besides that, modern version control platforms---such as GitHub and GitLab---are contributing to popularize code reviews by means of pull/merge requests.

However,  code review takes time and may introduce delays in the release of new versions. The reason is that it is a manual process that requires expertise in the codebase and careful inspection of textual diffs. In fact, diffs only provide a low level representation of code changes, which is based on added (+) and deleted lines of code (-). As a combined consequence of these two factors---manual inspections using line-based tools---software developers spend a considerable time performing code reviews. For example, Bosu and Carver~\cite{Bosu2013} and Bacchelli {\em et al.}~\cite{Bacchelli2018} report that, in specific contexts, the time spent in review tasks ranges from 3 to 6.4 hours per week, on average.

Particularly, current diff tools do not automatically detect refactorings in the inspected code change. For example, suppose a refactoring that moves a function \mcode{f()} from file \mcode{A} to file \mcode{B}. This move is represented in current state-of-the-practice diff tools as a set of lines deleted (-) from A and by a set of lines added (+) to \mcode{B}. Therefore, reviewers need first infer that these added/deleted lines indeed represent a move function. This is a not trivial inference when the change is complex, distributed over multiple files and when the refactoring is combined with other changes, such as bug fixes and the implementation of new features, as usual in practice \cite{fse2016danilo}. After inferring the refactoring, reviewers should also compare the code after and before the change, to review possible changes performed in the moved function. The reason is that refactorings are usually followed by minor edits in the code, particularly when we compare the changes at the level of commits.

\begin{figure*}[!ht]
\centering
\includegraphics[width=0.8\linewidth]{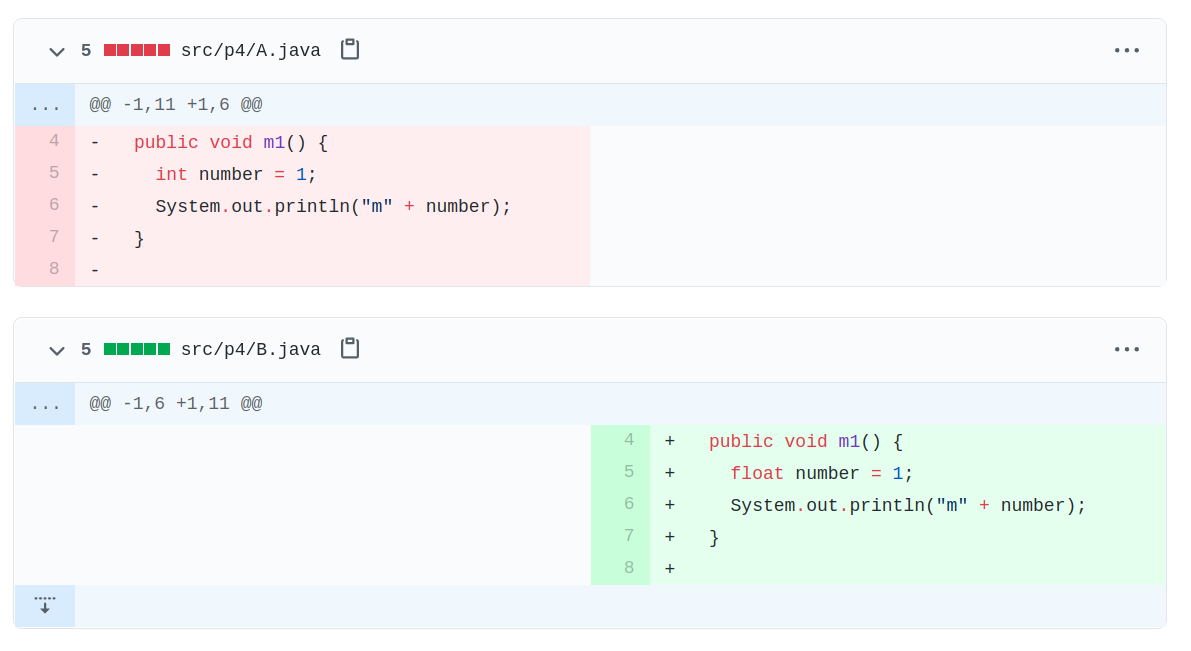}
\caption{Default diff including a Move Function refactoring (\mcode{m1} is moved from \mcode{A.java} to \mcode{B.java})}
\label{fig:move-github}
\end{figure*}

In this paper, we describe the key features and architecture of a tool for supporting {\em refactoring-aware code reviews}. Our goal is to alleviate the cognitive effort associated with code reviews, by automatically detecting refactoring operations included in pull requests. Besides supporting refactoring detection, the proposed tool---called RAID (or  Refactoring-aware and Intelligent Diffs)---seamlessly instruments current diff tools with information about refactorings. As a result, reviewers can easily inspect the changes performed in the refactoring code after the operation. Regarding its architecture, RAID is built on the top of recent and automatic refactoring tools. Particularly, RAID relies on RefDiff \cite{silva2020, silva2017}, which is the first tool that detects refactorings in other programming languages, besides Java. Finally, RAID operates with a low runtime overhead and it is fully integrated with state-of-the-practice continuous integration pipelines (GitHub Actions) and browsers (Google Chrome).

This paper is divided into three main sections:
\begin{enumerate}
    \item In Section \ref{sec:raid-nutshell}, we present RAID main features and Web-based interface. 
    \item In Section \ref{sec:raid-architecture}, we describe RAID internal architecture, including its integration with third-party tools, such as GitHub (Actions and pull requests) and browsers (Chrome).
    \item In Section \ref{sec:field-experiment}, we document the results and lessons learned in a field experiment with eight professional developers who used RAID during three months. We concluded that RAID can indeed reduce the cognitive effort required for detecting and reviewing refactorings. For example, the number of lines that need to be reviewed decreases from 14.5 to 2 lines (median values) in the case of move refactorings and from 113 to 55 lines (median values) in the case of extractions. We also collected the perceptions of the study participants about our tool, using to this purpose a post-experiment survey.
\end{enumerate}

RAID---our key contribution in this paper---is publicly available in this URL: \url{https://github.com/rodrigo-brito/refactoring-aware-diff}.

\section{RAID in a Nutshell}
\label{sec:raid-nutshell}

RAID is a tool for instrumenting textual diffs---particularly, the ones provided by GitHub---with information about refactorings. Typically, refactorings are represented in textual diffs as a sequence of removed lines in the left (-) and a sequence of added lines in the right. Therefore, code reviewers must infer by themselves whether this ``difference'' represents a refactoring operation, which requires an amount of cognitive effort. Figure \ref{fig:move-github} shows an example with a Move Function refactoring. In this case, method \mcode{m1} is moved from \mcode{A.java} (removed lines, in the left) to \mcode{B.java} (added lines, in the right).

\begin{figure}[!htb]
\centering
\includegraphics[width=0.9\linewidth,trim={0 0 21cm 0},clip]{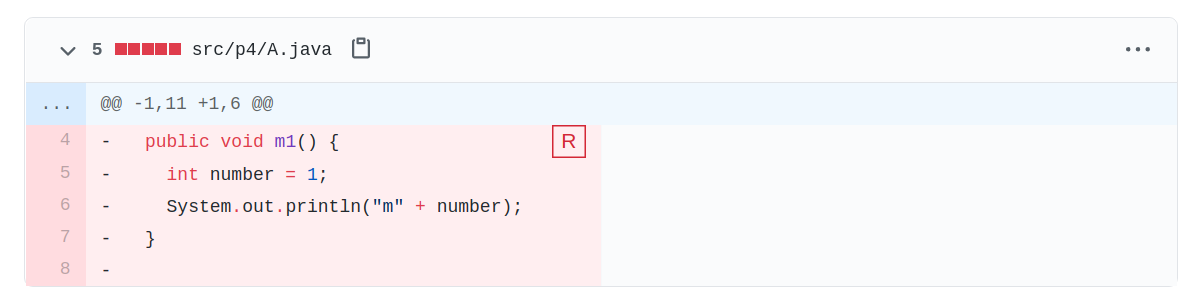}
\caption{Diff instrumented by RAID (an ``R'' button is added to the diff indicating the function is the part of a refactoring)}
\label{fig:r-annotation}
\end{figure}

Essentially, RAID preprocesses the default diff---provided by GitHub---in a seamless way and generates an enhanced diff visualization with explicit refactoring information. Figure \ref{fig:r-annotation} presents an example of refactoring annotation, in which RAID included an ``R'' button at the end of line four. This button is included in the refactoring source and destination, allowing for two-way navigation.

\begin{figure*}[!t]
\centering
\includegraphics[width=0.71\linewidth]{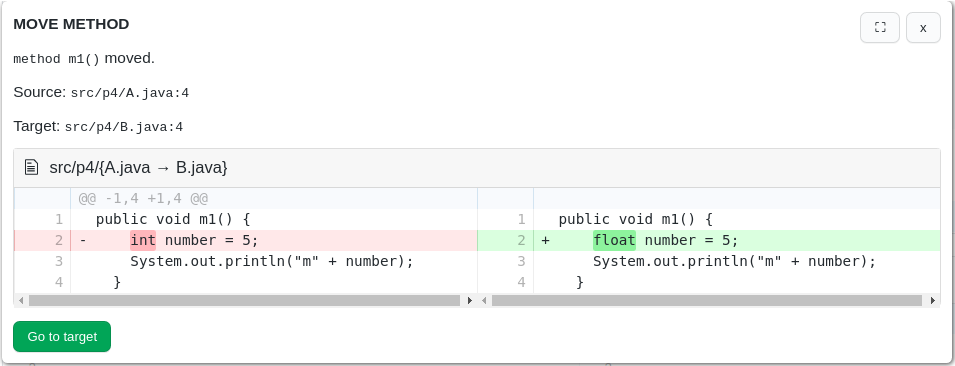}
\caption{Example of diff including a Move Function, as presented by RAID (this window is opened after clicking in the ``R'' button shown in Figure~\ref{fig:r-annotation})}
\label{fig:move-raid}
\end{figure*}

\begin{figure*}[h]
\centering
\includegraphics[width=0.71\linewidth]{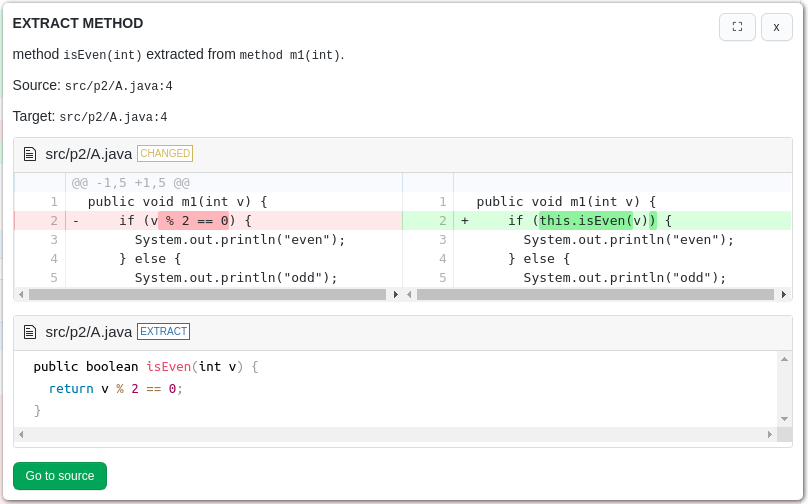}
\caption{Example of window documenting an Extract Function, as presented by RAID. We can see the lines changes in the original method (top) and also the code of the extracted method (bottom)}
\label{fig:extract-raid}
\end{figure*}

When the developer clicks in this button, a float window provided by RAID is displayed, as shown in Figure \ref{fig:move-raid}.  As we can see, this window includes detailed information about the detected refactoring. First, on the top, we have the following information about the operation: name (Move Function), short description (method \mcode{m1()} moved), source (\mcode{A.java}, line 4) and target file (B.java, line 4). Next, we can see a side by side view of the code before and after the move operation. Finally, in this internal diff, the lines changed in the moved code are highlighted, i.e., reviewers can easily review them. For the aforementioned reasons, we claim RAID reduces the cognitive effort required to review refactorings.

Extract Function is another refactoring that benefits from RAID features. As illustrated in Figure \ref{fig:extract-raid}, suppose a method \mcode{isEven} is extracted from a method \mcode{m1}. In this case, reviewers need to compare the code of the source method (\mcode{m1}) before and after the extraction. They also need to locate and review the code of the extracted method (\mcode{isEven}). As the reader can check in Figure \ref{fig:extract-raid}, RAID helps in both tasks. Particularly, the floating window provided by RAID includes, in the bottom, the code of the extracted method (\mcode{isEven}). Therefore, reviewers do not need to search for this method in the textual diff.

Besides move and extract refactorings, RAID provides information about other refactorings (when they are detected in a pull request). For example, in the case of pull up/push down refactorings, the information is similar to the one we described for move operations. In the case of Rename and Change Signature refactorings, RAID highlights the changes in the component's name and signatures. Table \ref{tbl:supported-refactorings} lists the refactorings detected by RefDiff/RAID.

\begin{table}[!h]
\centering
\caption{Refactorings detected by RefDiff/RAID}
\begin{tabularx}{0.9\linewidth}{@{}lX@{}}
\toprule
Language   & Refactorings                                                        \\ \midrule
Java       & Move, Extract Function, Inline Function, Rename, Change Signature, Pull Up, Push Down \\
JavaScript & Move, Extract Function, Inline Function, Rename, Change Signature, Pull Up, Push Down \\
C          & Move, Extract Function, Inline Function, Rename, Change Signature                     \\
Go         & Move, Extract Function, Inline Function, Rename, Change Signature                     \\ \bottomrule
\end{tabularx}
\label{tbl:supported-refactorings}
\end{table}

All windows provided by RAID include a ``Go to source'' button, which allows the reviewer to locate the source of the performed refactoring, thus reducing the effort to find the code in the default diff provided by GitHub.

\begin{figure}[!htb]
\centering
\includegraphics[width=\linewidth]{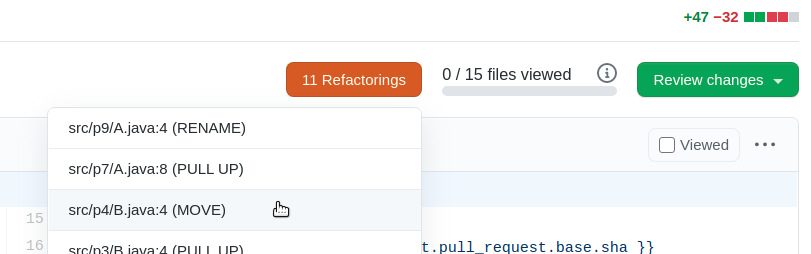}
\caption{RAID adds a button in the toolbar providing easy access to the list of refactorings in a pull request}
\label{fig:raid-index}
\end{figure}

Finally, RAID instruments the default diff's toolbar with a button that provides a list with all refactorings detected in the pull request. Figure \ref{fig:raid-index} shows an example of a pull request including 11 refactorings.

\section{RAID Architecture}
\label{sec:raid-architecture}

RAID architecture is composed by three components: (1) RAID GitHub Action (RGA), which is responsible for detecting and extracting refactorings; (2) RAID Chrome Extension (RCE), which is a browser extension that instruments GitHub's default diff pages with refactoring information; and (3) RAID Server, which provides a REST API for storing refactoring metadata and to connect the previous mentioned applications. 

\begin{figure}[!htb]
\centering
\includegraphics[width=0.7\linewidth]{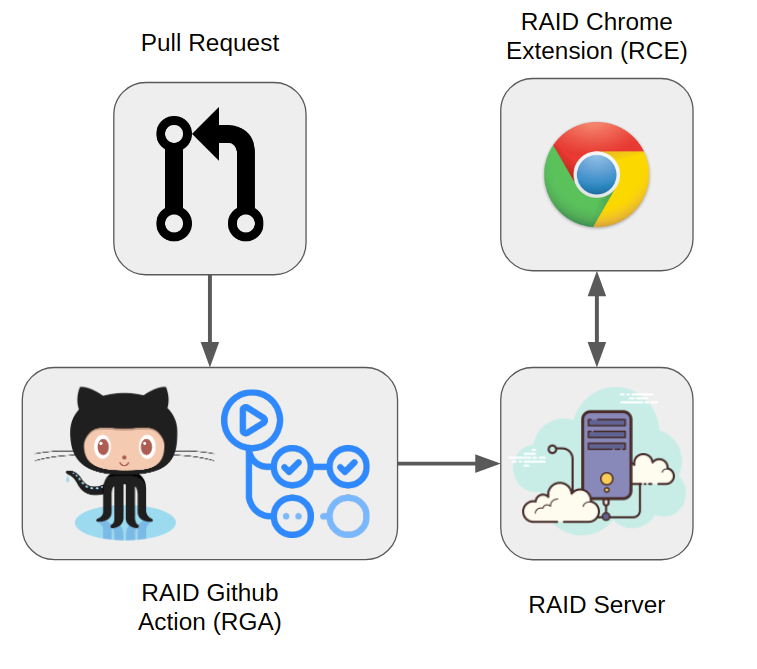}
\caption{RAID main components and execution flow}
\label{fig:raid-workflow}
\end{figure}

Figure \ref{fig:raid-workflow} presents RAID's workflow. First, when a developer submits a pull request, the RGA component is automatically called to detect the refactorings performed in the changed code. This data is then transmitted and stored by the RAID server. Finally, when a developer opens a diff page in his browser, RCE is automatically called to instrument the page with the refactoring information retrieved from the RAID server. 

In the following subsections, we describe each of these architectural components in details. 

\subsection{RAID GitHub Action (RGA)}

RAID GitHub Action component is responsible for analyzing the source code submitted by developers as part of a pull request. RGA is implemented in Java and it is executed in a Docker container. The main function of RGA is to detect the refactorings performed in a pull request. To this purpose,  RGA fully relies on a third-party tool: RefDiff \cite{silva2020, silva2017}, which is a multi-programming language refactoring detection tool. Specifically, RefDiff detects 13 types of refactorings in four programming languages: Java, JavaScript, C and Go. Consequently, RAID can be seamlessly used with any of such languages.

\begin{figure}[!htb]
\centering
\includegraphics[width=\linewidth]{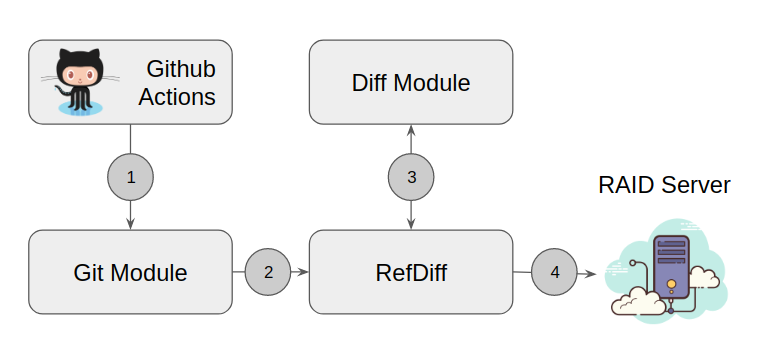}
\caption{RAID GitHub Action (RGA) modules and workflow}
\label{fig:modules}
\end{figure}

Figure \ref{fig:modules} shows RGA internal modules. As we mentioned, RGA is called after a pull request event is sent by GitHub Actions. In Step 1, RGA receives information about the pull request, such as branches, revisions, and commits. After that, the Git module loads the project and stores the commit history in a temporary directory.

At this point, it is important to remind that GitHub provides two distinct diffs for a given pull request: (1) the main diff, which compares changes performed in the last commit of the pull request with the repository HEAD; and (2) individual commit diffs, which compare the changes performed in a given commit $C_i$ with its parent commit. Therefore, RefDiff is called by RAID to extract the refactorings included in each of these diffs. To be clearer, assume a pull request with three commits: $C_1$, $C_2$, and $C_3$. In this case, RAID calls RefDiff four times, with the following pairs of commits: ($C_3$, HEAD), ($C_3$, $C_2$), ($C_2$, $C_1$), and ($C_1$, HEAD).

Next, the Diff Module computes the internal diffs that are showed in RAID's floating windows (see Figures \ref{fig:move-raid} and \ref{fig:extract-raid}). For example, in the case of a Move Function refactoring, RAID highlights the changes performed in the moved method (since move refactorings might include minor changes in the code after its movement \cite{silva2020, tsantalis2020}).  Finally, RGA transmits the refactorings data to the RAID Server.

\subsection{RAID Chrome Extension (RCE)}

The RAID Chrome Extension (RCE) presents the refactoring data provided by the RAID server. RCE is implemented in JavaScript (\url{~}1.1 KLOC) and it currently supports Google Chrome browser. The extension is automatically called by Chrome after a valid GitHub's pull request URL is opened by the browser. This URL includes information about the pull request, which is used by RCE to contact the RAID server and to obtain data about possible refactorings.

For each refactoring, RCE transparently modifies the DOM (Document Object Model) of the default visualization provided by GitHub and inserts a button, identified by the letter ``R'', at the end of the line where the refactoring was identified, as we already illustrated in Figure  \ref{fig:r-annotation}. The RCE module is also responsible for displaying the information of refactoring activities, as presented in Figure \ref{fig:move-raid}. Finally, RCE assists the reviewer to navigate between the refactored code elements.

\subsection{RAID Server}

The RAID Server acts as a bridge between the RCA and RCE components, i.e., it receives the refactoring information after each RGA execution and sends the refactorings to RCE when requested. To this purpose, the server provides a REST API implemented in the Go language. Internally, the server stores the refactoring data in a non-relational database at Google Firebase.\footnote{\url{https://firebase.google.com}} For open-source projects, RAID provides a public API. For private repositories, the server supports a custom credentials configuration to control user access.

\section{Field Experiment}
\label{sec:field-experiment}

\subsection{Methodology}

To evaluate RAID, we relied on a field experiment. More specifically, we obtained permission to include the tool in the development workflow of a medium-sized technology company that develops software for musical products. In this way, two development teams (with 5 and 3 developers) used the tool during three months, from June to September 2020. Table \ref{tbl:projects-data} describes basic information about the software projects and the pull requests performed during the experiment. All four projects are implemented in the Go programming language.

\begin{table}[!htb]
\centering
\caption{Number of pull requests, commits, and lines of code by project}
\label{tbl:projects-data}
\begin{tabular}{@{}lrrrr@{}}
\toprule
Name      & Pull Requests & Commits & LOC & Team \\ \midrule
Project A & 94            & 102     & 370K    & 1 \\
Project B & 86            & 89      & 196K    & 1 \\
Project C & 54            & 65      & 34K     & 1 \\
Project D & 91            & 118     & 154K    & 2 \\ \midrule
All       & 325           & 374     & 754K     &\\ \bottomrule
\end{tabular}
\end{table}

In this company, all development teams use GitHub for version control. Developers also use pull requests followed by code reviews to integrate new code in the projects' repository. First, developers perform modifications in private forks and submit pull requests to the main repository. Then, a CI server is used to run linters and a suite of unit tests. Finally, code review is carried out before integrating the change in the mainline. The code needs to be approved by two team members who did not work in the change.

Before starting the experiment, the participants received instructions and training on our tool (\url{~}30 minutes), followed by one week of warm-up and to get used with RAID. In addition, the first author of this paper provided support during the study period to clarify questions.

For the experiment, we instrumented RAID to collect all refactorings performed during the study period and also the user events. These events include the clicks in the ``R'' and ``Go to source'' buttons, and the time interval between opening and closing the floating windows with refactoring descriptions.

\begin{figure}[!htb]
\centering
\includegraphics[width=\linewidth]{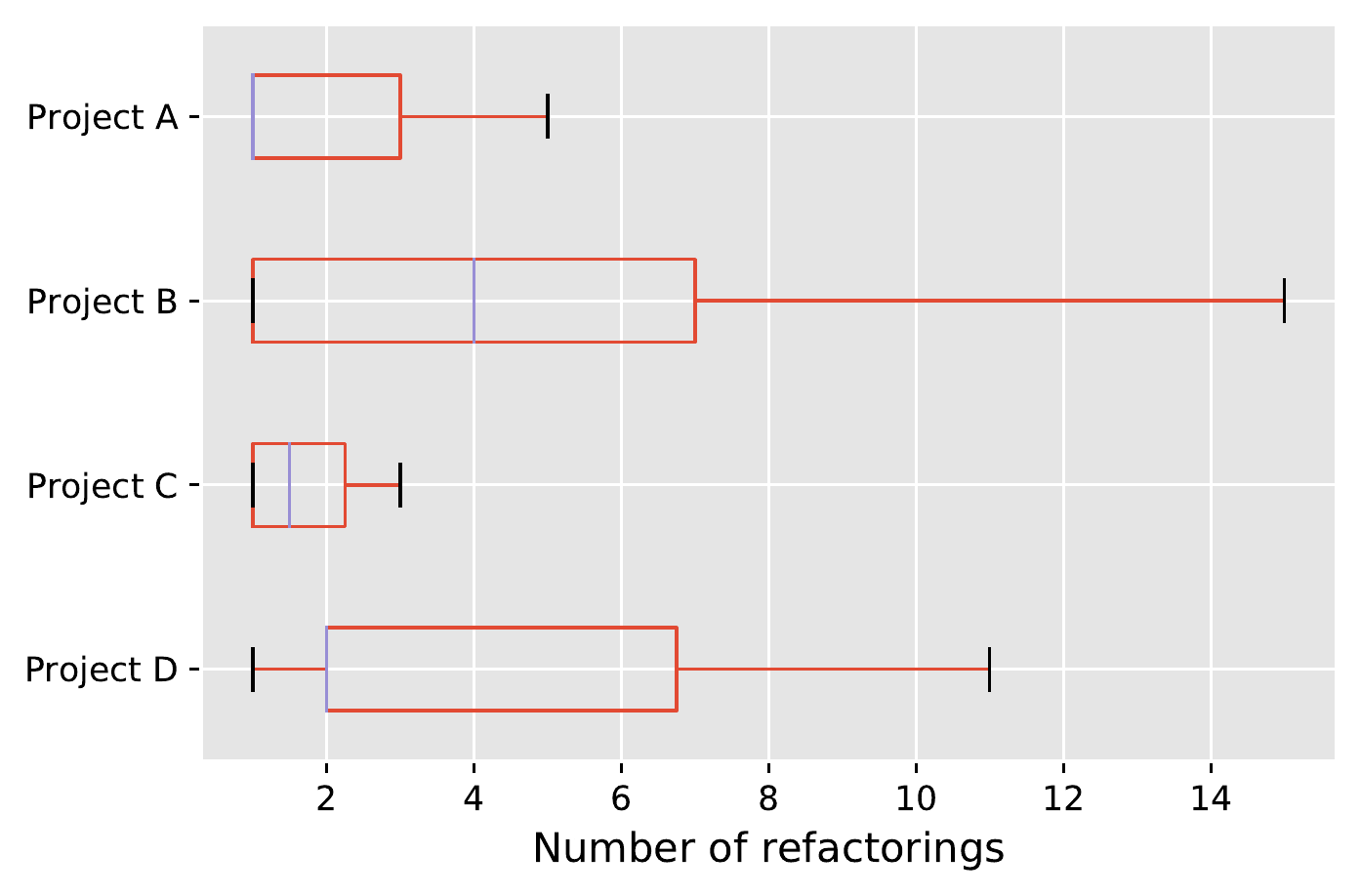}
\caption{Number of refactorings per pull request}
\label{figure:refactorings-pr}
\end{figure}

During the experiment, 325 pull requests were created. Out of them, 84 pull requests (26\%) included refactoring activities. Figure~\ref{figure:refactorings-pr} presents the distribution of refactorings for these 84 pull requests. The project with the highest number of refactorings per pull request is project B (median of 4 refactorings per PR), followed by project D (median of 2).

\begin{figure}[!htb]
\centering
\includegraphics[width=\linewidth]{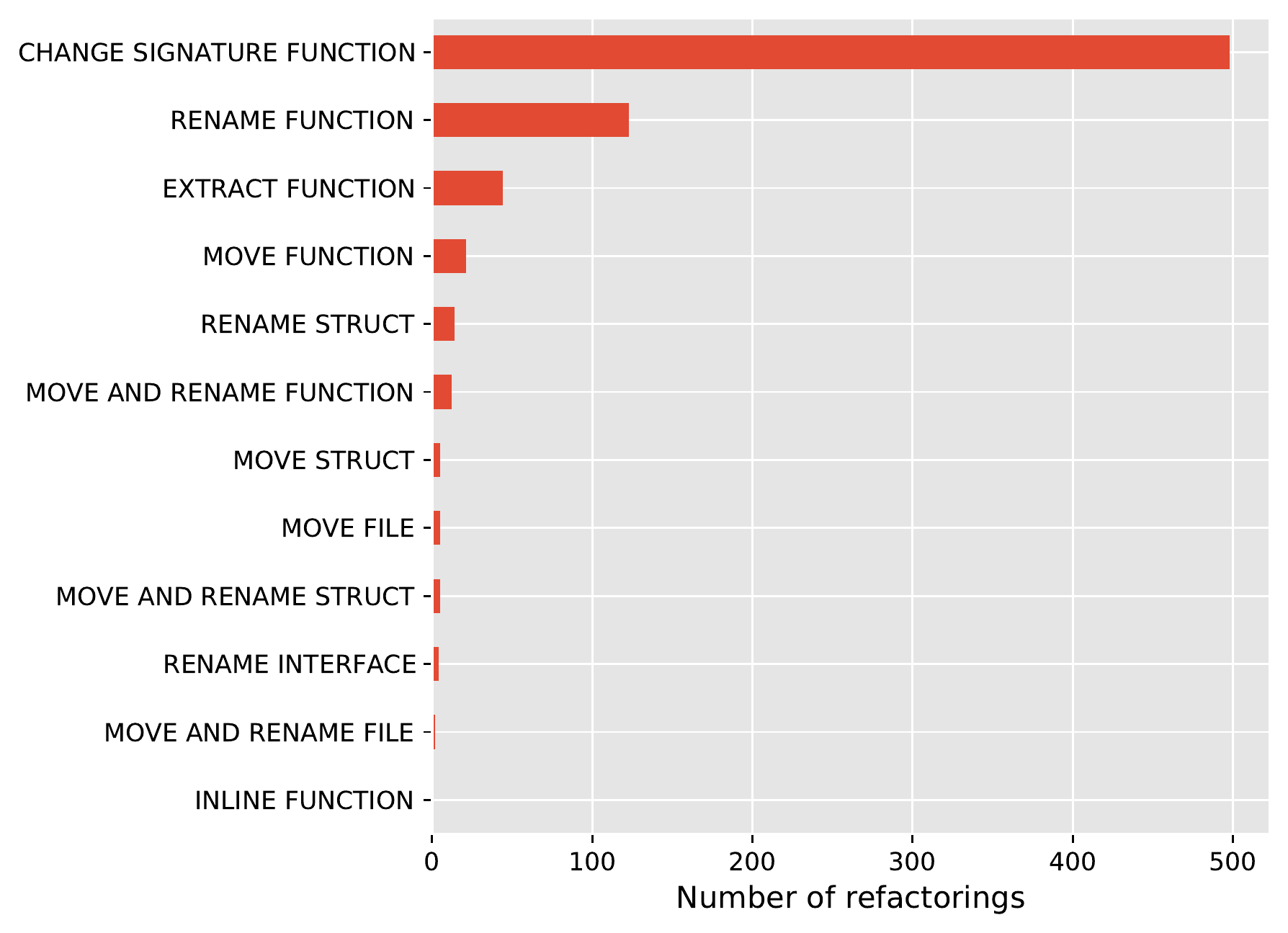}
\caption{Most frequent refactorings}
\label{figure:refactoring}
\end{figure}

In total, we collected data about 685 refactorings. As we can see in Figure~\ref{figure:refactoring}, the top-3 refactorings are Change Signature (498 occurrences, 73\%), Rename Function (123 occurrences, 18\%), and Extract Function (44 occurrences, 6\%).

After the experiment, we sent a short survey to the participants asking their perceptions about RAID.  In this survey, we asked three questions:
\begin{itemize}
    \item What are the key benefits of RAID?
    \item What are the difficulties you experienced using RAID?
    \item Any additional comments or suggestions?
\end{itemize}

\noindent We received responses from all 8 participants.

\subsection{Research Questions}

We used the data collected in the field experiment to answer four research questions:

\vspace{0.2cm}

\noindent \textit{RQ1: What is the runtime overhead introduced by RAID?} \vspace{0.1cm}

\noindent We started with this question since it is important to check whether RAID introduces (or not) a delay in the code review process. In other words, after submitting a pull request, it is important that the information provided by RAID becomes available to code reviewers as soon as possible.

\vspace{0.2cm}

\noindent \textit{RQ2: How did developers use RAID during code review?} \vspace{0.1cm}

\noindent With this second RQ, we characterize the usage of RAID during the three months of our experiment. For example, we report and analyze the collected usage data, including most frequent refactorings with user events, reviewing time, and UI events performed by users on RAID's web-based interface.

\vspace{0.2cm}

\noindent \textit{RQ3: How much cognitive effort is reduced with RAID?} \vspace{0.1cm}

\noindent Although this is the key goal of RAID, it is not trivial to estimate the amount of cognitive effort that is saved when RAID is used to support code review tasks, particularly in the case of an experiment performed in the context of a real software company. For this reason, we estimate this effort using two proxies: Diff Code Churn (DCC), which is the number of lines needed to represent a code change in a textual diff and the distance of the code moved by a refactoring (if any).

\vspace{0.2cm}

\noindent \textit{RQ4: How did developers perceive RAID?} \vspace{0.1cm}

\noindent  In this last research question, we report the perceptions of the participants about our tool, as commented by them in the post-experiment survey.

\subsection{Experiment Results}

\noindent \textit{RQ1: What is the runtime overhead introduced by RAID?} \vspace{0.1cm}

\noindent Regarding the execution time required by RAID to identify the refactorings, Projects A, C, and D have a median time lower than 40 seconds, while Project B has a median of 95.7 seconds. However, as previously presented in Figure \ref{figure:refactorings-pr}, project B has the highest median number of refactorings per pull request. Moreover, most high runtime results observed in Project B refer to atypical changes performed in the code base, such as updating third party code. However, since the CI server runs RAID in parallel with other time-consuming activities such as unit tests, even a median time of 95 seconds is acceptable in practical code review scenarios.

\begin{figure}[!htb]
\centering
\includegraphics[width=\linewidth]{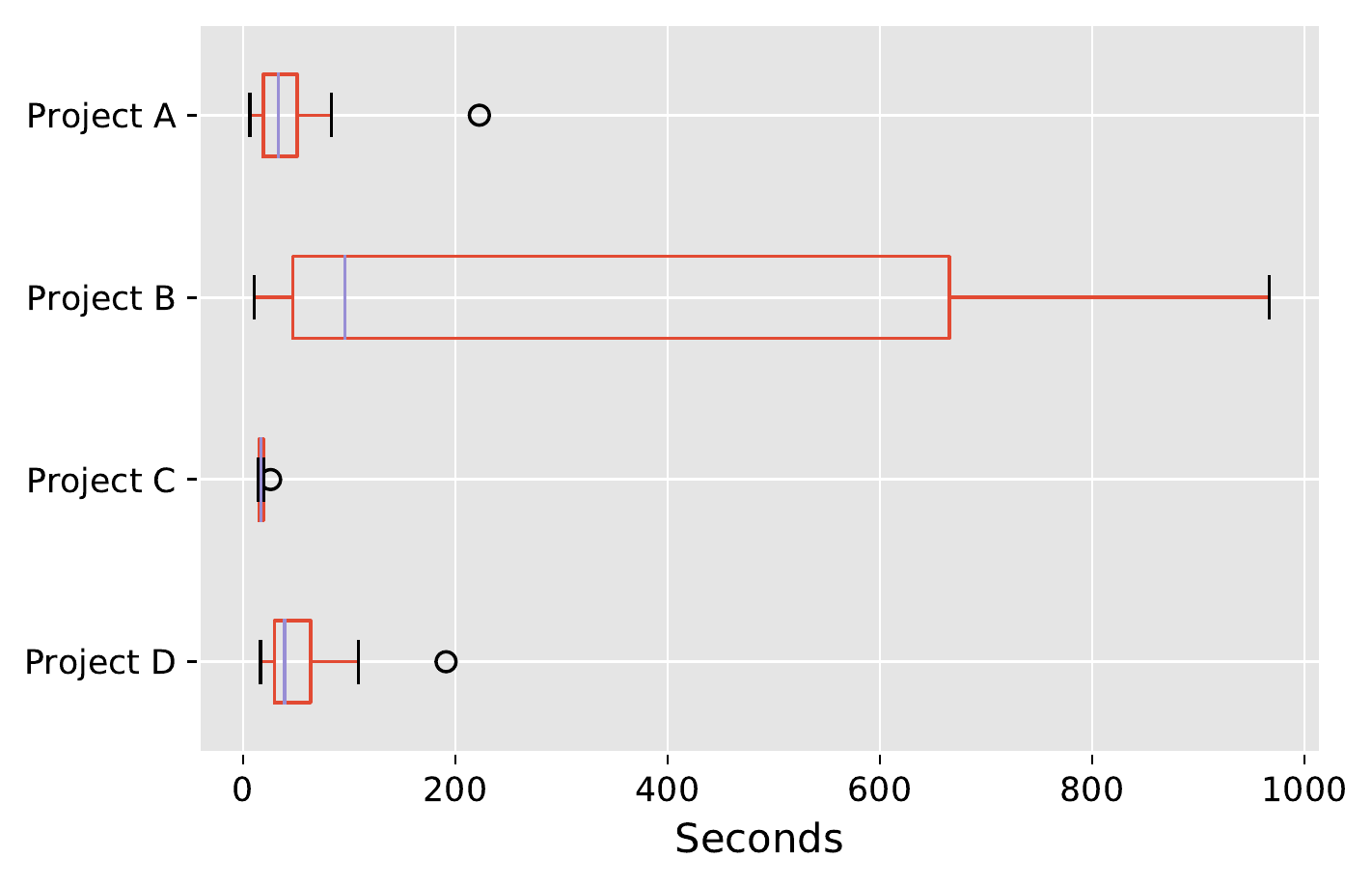}
\caption{Execution time per project}
\label{figure:execution-time}
\end{figure}

\begin{tcolorbox}[left=0mm,right=0mm,boxrule=0.25mm,colback=gray!5!white]
	\vspace{-0.2cm}
	{\em Summary:} Typically, the information provided by RAID is available to code reviewers in less than 2 minutes after submitting the pull request. Therefore, in practical terms, RAID does not delay the start of code reviews.
\end{tcolorbox}

\noindent \textit{RQ2: How did developers use RAID during code review?} \vspace{0.1cm}

\begin{figure}[!htb]
\centering
\includegraphics[width=\linewidth]{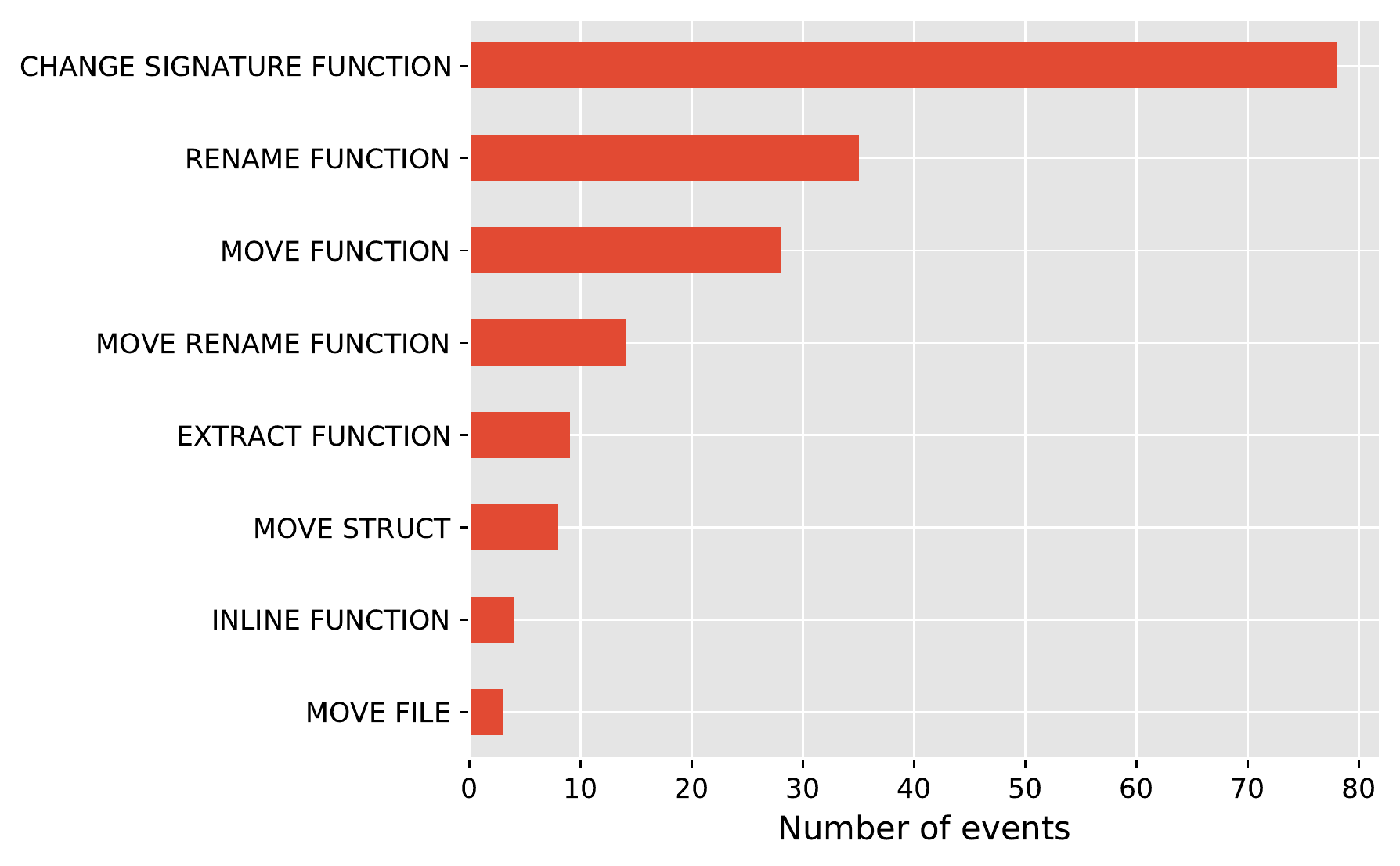}
\caption{Most common refactorings with clicks in the ``R'' buttons}
\label{figure:events-by-refactoring}
\end{figure}

\noindent Figure \ref{figure:events-by-refactoring} presents the most common refactorings with user events, i.e, refactoring operations performed in pull requests that were inspected by code reviewers through RAID, by clicking in the ``R'' buttons. As we can notice, most cases refer to refactorings performed over functions. Specifically, the  top-3 operations comprise Change Signature Function (78 occurrences, 43.6\%), Extract Function (35 occurrences, 19.6\%),  and Move Function (28 occurrences, 15.7\%). 

\begin{figure}[!htb]
\centering
\includegraphics[width=\linewidth]{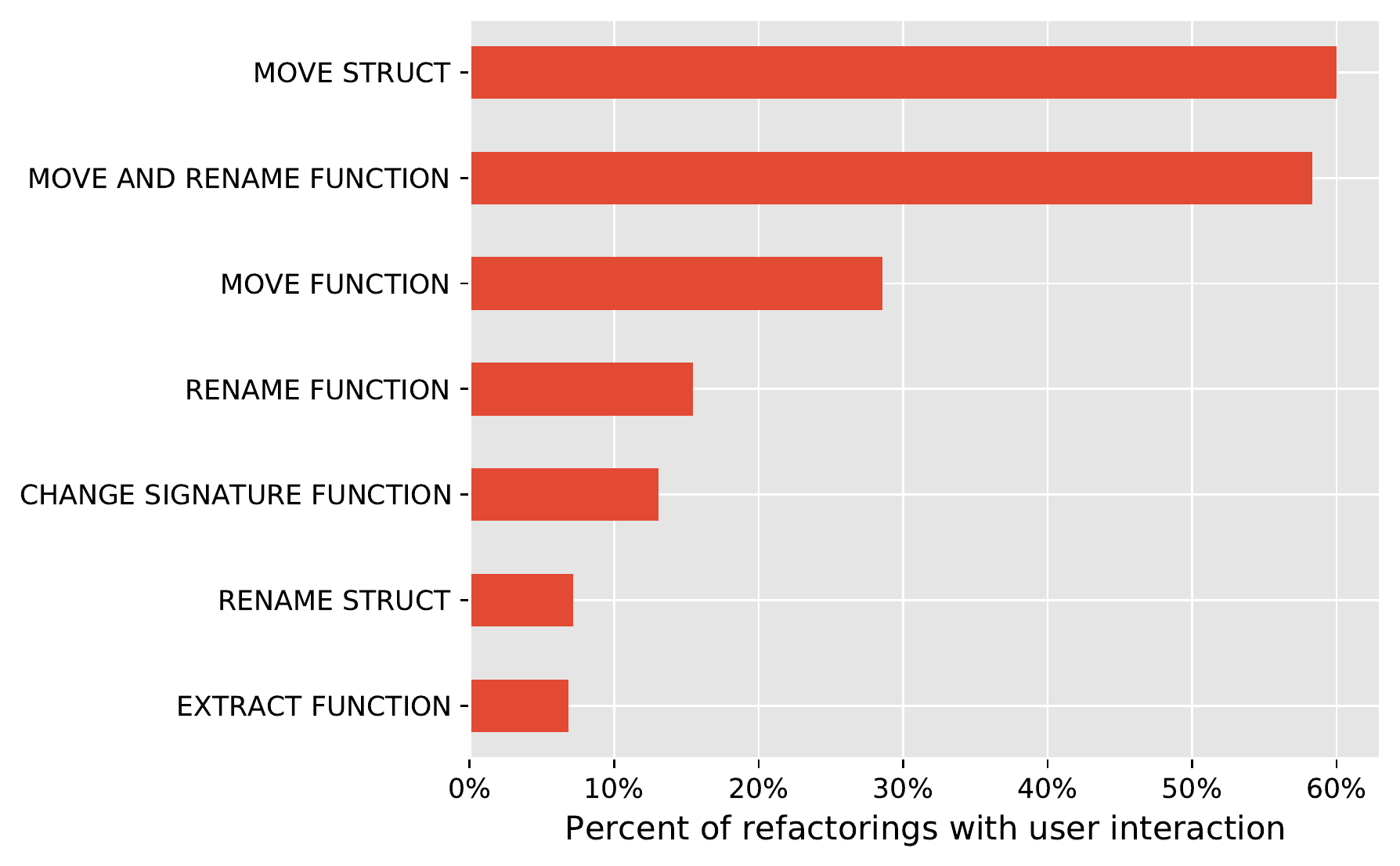}
\caption{Ratio of clicks in ``R" buttons, i.e., number of times the button received a click divided by number of times it was shown}
\label{figure:events-percent}
\end{figure}

We also analyzed the percentage of clicks in the ``R'' buttons.
Assume a button is presented $x$ times. As a result, it
received $y$ clicks by code reviewers. The percentage represented
in Figure \ref{figure:events-percent} is the ration $y / x$. However, in this figure
we only considered refactorings that appeared at least
10 times, i.e., $x \geq 10$. The reason is that the described ratio
is not representative in the case of rare refactoring
operations (e.g., 100\% is a very different result depending on whether the button appeared 1 or 100 times). 

As we can see in Figure \ref{figure:events-percent}, the top-3 most clicked refactorings (in percentage values) represent move operations, i.e., Move Struct (60\%), Move and Rename Function (58\%), and Move Function (29\%). Interestingly, Change
Signature is the most common refactoring (in absolute terms, Figure 11).
However, only 13\% of the ``R" buttons associated to particular
instances of this refactoring received clicks. A probable
reason is that Change Signature is a relatively simple
refactoring, when compared with move ones. 


\begin{figure}[!htb]
\centering
\includegraphics[width=0.8\linewidth]{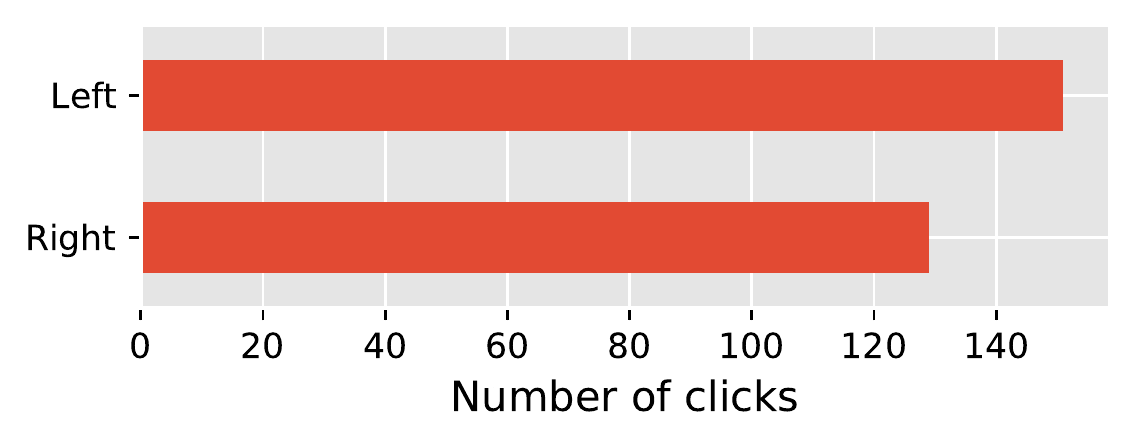}
\caption{Number of clicks on the left and right buttons provided by RAID}
\label{figure:events-by-side}
\end{figure}

GitHub also allows users to split the screen for a side-by-side comparison of the changes in a textual diff, i.e., the original version is presented on the left side and the modifications are presented on the right side (see an example in Figure \ref{fig:move-github}). In this case, when a refactoring is detected, RAID adds an ``R'' button on both sides of the diff. For example, when a method is moved, on the left side a ``R'' appears on top of the deleted lines; and an identical button is added on the right side, on top of the new position of the method. Therefore, when reviewing refactorings, developers can click on both buttons. We instrumented RAID to collect the number of such clicks and the results are presented in Figure \ref{figure:events-by-side}. As we can see, the click events are almost equally distributed between the two sides of a diff, with a minor preference for clicks on the left side (53.9\%).

\begin{figure}[!htb]
\centering
\includegraphics[width=\linewidth]{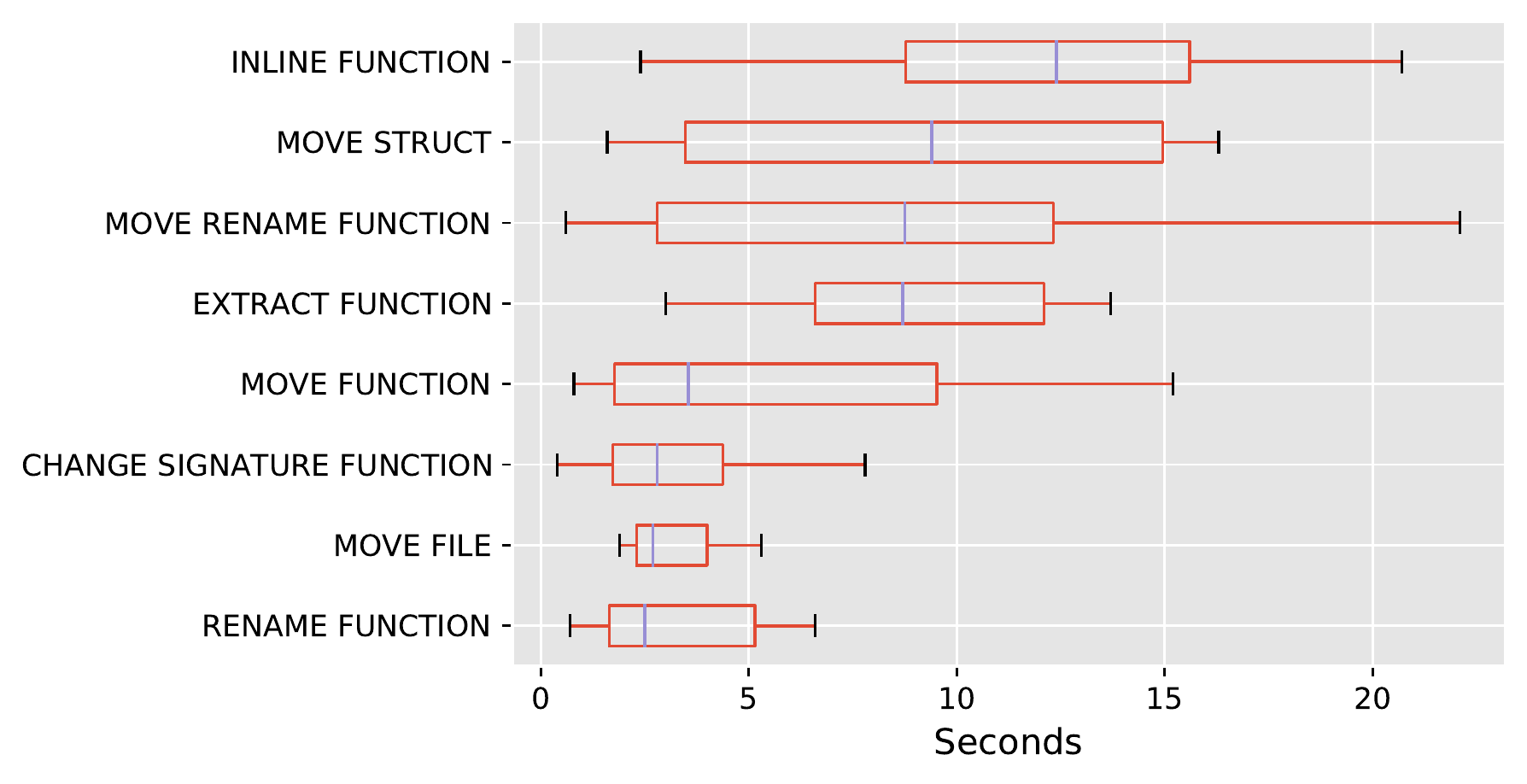}
\caption{Time spent reviewing the information provided by RAID (per refactoring operation)}
\label{figure:inspection-time}
\end{figure}

Finally, we computed the time spent by code reviewers when inspecting the information provided by RAID in the floating windows. The results are presented in Figure \ref{figure:inspection-time}. Each distribution refers to the amount of time the reviewer kept the window opened. Considering the median of the distributions, the top-5 refactorings that required more reviewing time are Inline Function (12.4 sec), Move Struct (9.4 sec), Move and Rename Function (8.7 sec), Extract Function (8.7 sec), and Move Function (3.55 sec). Therefore, we can see that refactorings that involve moving code demand more time to consume the information provided by RAID. 

By contrast, refactorings such as Change Signature Function, Move File, and Rename Function have low inspection time, with a median of less than 3 seconds. Indeed, both Change Signature and Rename Function are simple operations, where RAID does not provide improvements to the traditional diff, which probably explains their lower time.

\begin{tcolorbox}[left=0mm,right=0mm,boxrule=0.25mm,colback=gray!5!white]
	\vspace{-0.2cm}
	{\em Summary:}  (1) in relative terms, reviewers relied on RAID to obtain information mostly for Move Struct, Move and Rename Function, and Move Function; (2) reviewers used both ``R'' buttons---available on the left and right side of the original diff---to request refactoring information provided by RAID; (3) on the median, reviewers spent from 2.5 seconds (Rename Function) to 12.4 seconds (Inline Function) to interpret the refactoring information provided by RAID.

	\vspace{-0.2cm}
\end{tcolorbox}

\noindent \textit{RQ3: How much cognitive effort is reduced with RAID?} \vspace{0.1cm}

\noindent In this RQ, we provide indicators of the impact on cognitive effort achieved with RAID. For this purpose, we concentrate on two refactorings: Move and Extract Function, since they are more complex than simple refactorings, such as a rename.

\vspace{0.2cm}

\noindent \textit{Move Refactorings.} Out of 32 move refactorings performed during the experiment, 11 operations move code between files located in different packages and 21 operations are performed in files from the same package, Among these, 9 operations are performed within the same file but between internal \mcode{structs}. Therefore, most of the move refactorings impact distinct files (71.8\%). We claim it is more difficult to detect and review such moves when using traditional diffs. The reason is that the default visualization lacks information about the refactored files. On the other hand, with RAID developers can easily access this information, as illustrated in Figure \ref{fig:move-raid}.

We also found nine move refactorings that were performed in the same file. In Go, an internal move occurs when a function is moved to a new \mcode{struct} or moved between existing \mcode{structs} located in the same file. In this case, we computed the distance of the moved code, i.e., suppose a move from line $x$ to line $y$, the distance in this case is $| y - x |$. The rationale is that the higher this distance, the higher the effort for a reviewer to detect and inspect the move, assuming she is using a non-refactoring aware diff tool. By contrast, with RAID, this effort does not exist since the tool automatically detects the refactoring and provides buttons to navigate from its source to target lines (and back). 

Table \ref{tbl:move-distance} shows the values of the distance metric for the nine move refactorings restricted to the same file. As we can see, eight refactorings moved the code at least 40 lines above or below the original position. In one case, this distance is zero due to a coincidence, when multiple refactoring operations ended up aligning the diffs after the move. Therefore, the effort to locate the moved code---even when it is located in the same file---is not trivial, assuming code reviewers rely on traditional diffs. For example, a distance of 41 lines means the function is one page above or below the code under the reviewer focus in the browser.

\begin{table}[!htb]
\caption{Distance of Move Function refactorings, i.e., line of the moved code after the operation minus line of the code before the operation}
 \label{tbl:move-distance}
\begin{tabular}{@{}lrrrrrrrrr@{}}
\toprule
Refactoring & 1 & 2  & 3  & 4  & 5  & 6   & 7   & 8   & 9    \\ \midrule
Distance    & 0 & 40 & 40 & 40 & 41 & 466 & 466 & 474 & 4,870 \\ \bottomrule
\end{tabular}
\end{table}

\begin{figure}[!htb]
\centering
\includegraphics[width=\linewidth]{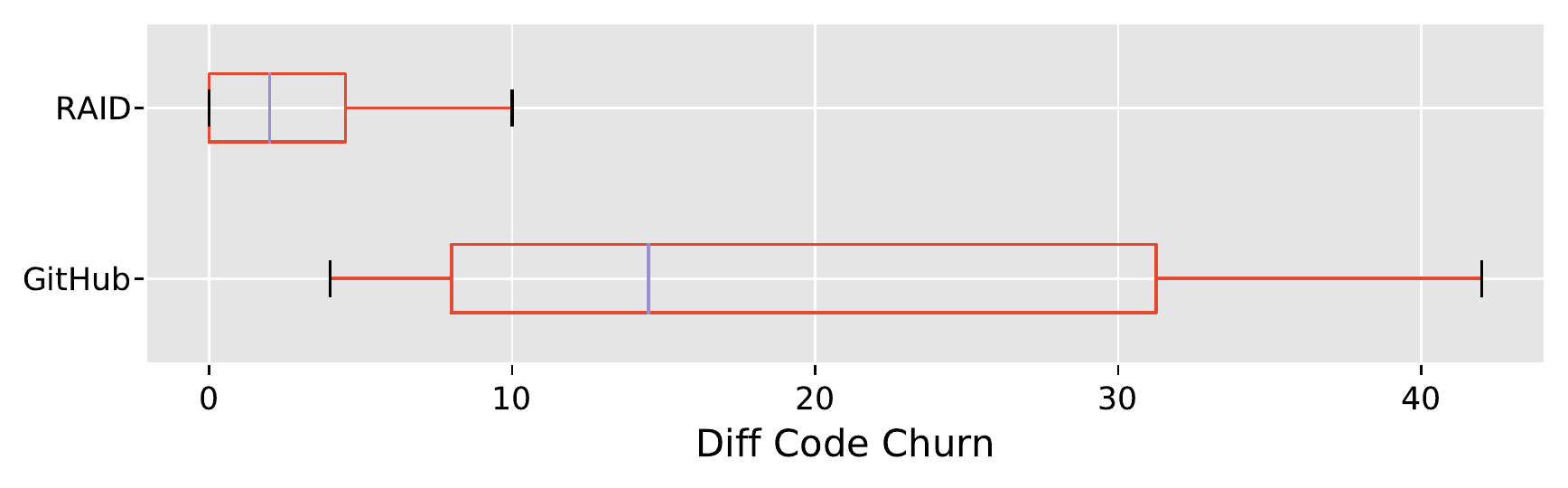}
\caption{Diff Code Churn (DCC) when reviewing Move Function refactorings, when using RAID and GitHub diff}
\label{figure:proxy-move}
\end{figure}

We also computed the Diff Code Churn (DCC) of the move refactorings. We refer  to Diff Code Churn as the number of added (+) and deleted (-) lines showed in the diff interface to represent a refactoring. Considering the 32 move refactoring performed in the experiment, DCC is 14.5 lines of code (median values), as presented in Figure \ref{figure:proxy-move}. On the other hand, when using RAID to review these moves, DCC is reduced to just 2 lines of code (median values). The reason is clear: RAID aligns the moved lines of code and only shows the minor changes performed in the code after the move (using + and - lines). Examples are presented in Figures \ref{fig:move-github} and \ref{fig:move-raid}. In Figure \ref{fig:move-github}, we have a move represented in a traditional diff, where DCC is 10 lines (5 added plus 5 deleted lines). In Figure \ref{fig:move-raid}, we have the same move, as detected and presented by RAID. As we can see, DCC is just 2 lines.

\vspace{0.2cm}

\begin{tcolorbox}[left=0mm,right=0mm,boxrule=0.25mm,colback=gray!5!white]
	\vspace{-0.2cm}
	{\em Summary:} (1) 71.8\% of the Move Function refactorings are inter-files, when RAID clearly outperforms non-refactoring aware diffs; (2) In the case of intra-file moves, the code is moved 41 lines after or below its original position (median results); (3) Diff Code Churn reduces from 14.5 lines to only 2 lines (median results), when moves are reviewed using RAID.
	\vspace{-0.2cm}
\end{tcolorbox}

\begin{figure}[!htb]
\centering
\includegraphics[width=\linewidth]{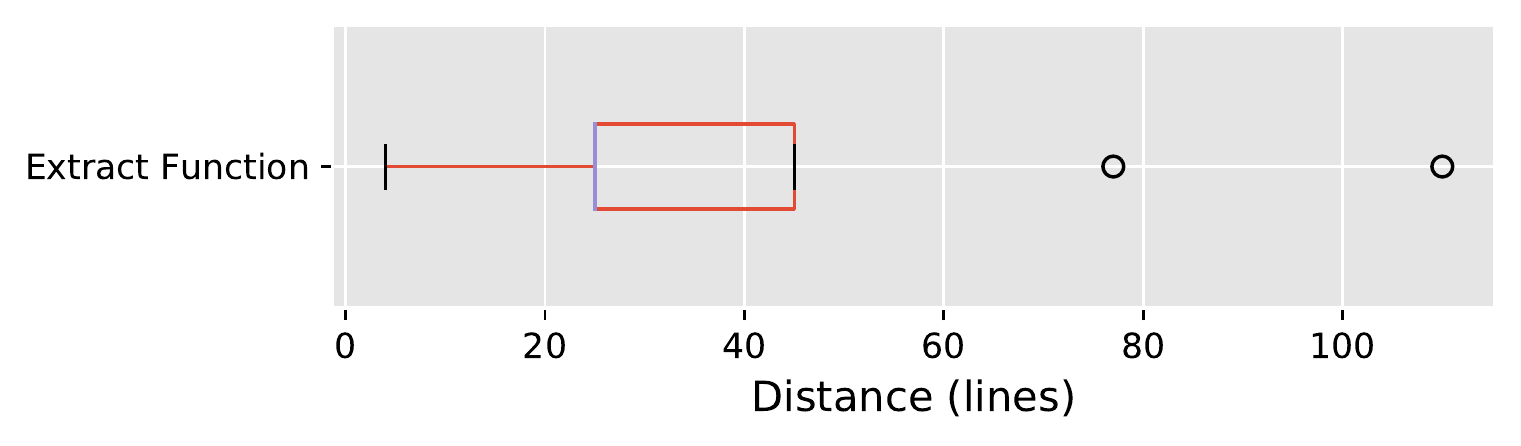}
\caption{Distance of Extract Method refactorings, i.e., initial line of the extracted method minus initial line of the source method, in absolute terms}
\label{figure:proxy-extract-lines}
\end{figure}

\noindent \textit{Extract Refactorings.} We found 44 occurrences of Extract Function. First, we compared the distance between the first line of the source code and the first line of the extracted code. Figure \ref {figure:proxy-extract-lines} shows the distribution of the distance values. We can see that the extracted code is implemented 25 lines after or before the source method (median values). Therefore, these numbers suggest that it is probably not difficult to locate the method. On the other side, before locating the extracted method, reviewers should infer whether the diff contains an Extract Function, which might not be trivial.

\begin{figure}[!htb]
\centering
\includegraphics[width=\linewidth]{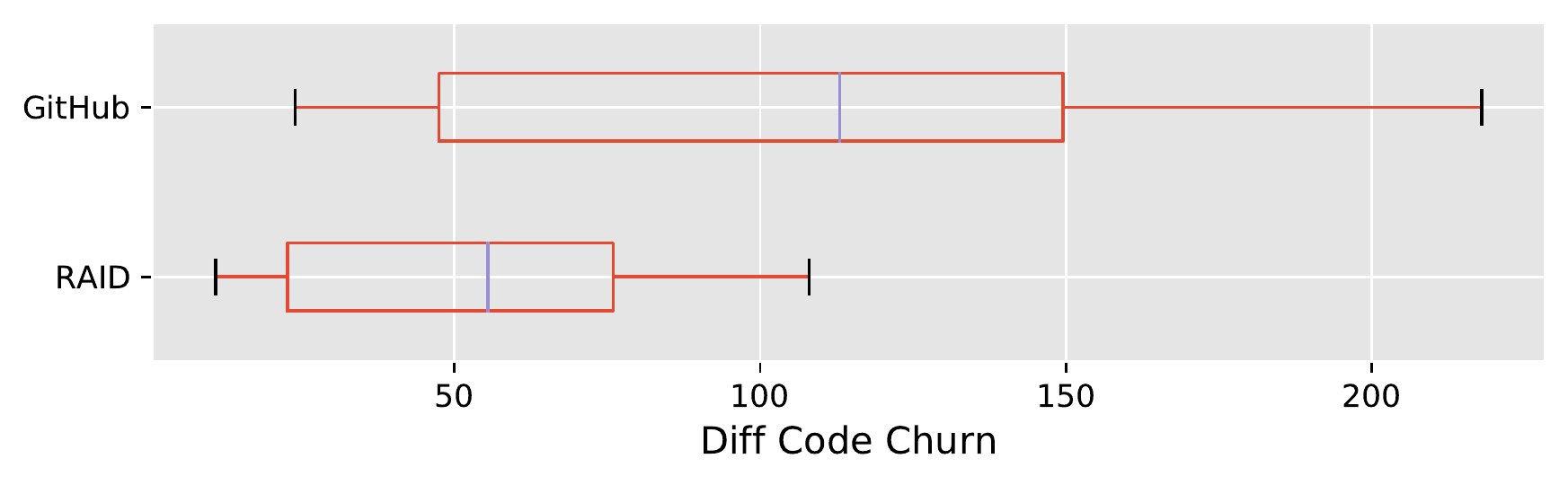}
\caption{Diff Code Churn when reviewing Extract Function refactorings, when using GitHub Diff and RAID}
\label{figure:proxy-extract-vcc}
\end{figure}

We also computed the Diff Code Churn for extract refactorings. Figure \ref{figure:proxy-extract-vcc} shows the distribution of DCC values, when using non-refactoring aware diffs and RAID. The median values decrease from 113 lines (when using GitHub's diff) to 55 lines when using RAID. The reason is that RAID reduces the number of lines to be analyzed, due to the precise identification of the extracted code. Figure \ref{fig:move-raid} shows an example, when RAID highlighted the single line of code changed in the extracted method (in this case, just the variable type changed from \mcode{int} to \mcode{float}).  

\begin{tcolorbox}[left=0mm,right=0mm,boxrule=0.25mm,colback=gray!5!white]
	\vspace{-0.2cm}
	{\em Summary:} (1) Methods are extracted 25 lines before or after the source method; (2) Diff Code Churn decreases from 113 lines to 55 lines when using RAID to review Extract Function refactorings.
\end{tcolorbox}

\noindent \textit{RQ4: How did developers perceive RAID?} \vspace{0.1cm}

\noindent In this section, we report the results of the post-experiment survey.

\vspace{0.2cm}

\noindent \textit{RAID Benefits.} All participants mentioned that RAID represents an improvement regarding GitHub's traditional diff. As examples, we received the following answers.

\vspace{0.2cm}

\noindent \textit{RAID highlights large migrations of code, easing code review when the snippet was just moved without major changes. (S1)}

\vspace{0.2cm}

\noindent \textit{RAID allows a more efficient code review with a realistic diff, which shows the real changes that occurred in the code. (S2)}

\vspace{0.2cm}

\noindent \textit{The tool makes the refactor recognition process instantaneous, and shows a more targeted and clean diff, facilitating and accelerating the understanding of the code.  (S8)}

\vspace{0.2cm}

\noindent \textit{Limitations.} We received five responses pointing out limitations in RAID. For example, one reviewer reported that RAID did not identify refactorings in specific cases.

\vspace{0.2cm}

\noindent \textit{I particularly experienced some bugs where RAID did not detect possible refactors. (S7)}

\vspace{0.2cm}

Indeed, RAID is based on RefDiff, which has 92\% of precision and 80\% of recall for Go~\cite{brito2020}. Particularly, the tool might miss refactorings when developers, for example, make considerable changes in the code after the operation. In these cases, the tool might not detect a refactoring although a reviewer might still consider it as such.

Finally, two reviewers reported the limitation of browser support. Currently, RAID supports only Google Chrome. As future work, we intend to extend the tool to other browsers, such as Safari and Mozilla Firefox.

\section{Threats to Validity}
\label{sec:threats-validity}

\noindent \textit{External Validity.} In our field experiment, we evaluated four Go language projects from a medium-sized company. The participants used RAID during three months. Therefore, on the one hand, we strived to conduct a real experiment, with real-world systems and professional developers. On the other hand, we acknowledge that our results may change if we consider more projects, more developers, a longer time box, and other programming languages.

\vspace{0.2cm}

\noindent \textit{Construct Validity.} First, the refactoring detection module relies on RefDiff~\cite{silva2020}, which has a precision of 92\% and recall of 80\%, in the case of Go projects~\cite{brito2020}. In this way, RAID visualizations might miss refactorings (false negatives) or may incorrectly detect refactorings (false positives). Indeed, the presence of false negatives was mentioned by one of the experiment participants. However, we highlight that RefDiff accuracy is compatible with the state-of-the art in refactoring detection tools~\cite{tsantalis2020, tsantalis2018}. Furthermore, to mitigate the impact of this threat on the field experiments results, the first author of this paper manually evaluated each of the refactorings used to compute the proxy metrics presented and discussed when answering RQ3. In total, we found 10 false positives (13.2\%) among moves and extracts.

\vspace{0.2cm}

\noindent \textit{Internal Validity.} A first possible internal threat to validity can be found in RQ1, where the size of pull requests (i.e., the number of modified lines) might have an influence on the execution time. In our third research question, we proposed  two proxies to estimate the amount of cognitive effort reduced by RAID: Diff Code Churn (DCC) and distance of move refactorings. On the one hand, claim that such proxies have a positive correlation with the cognitive effort required in code reviews. On the other hand, they did not allow us to provide a clear measure of the effort reduction. Therefore, further studies---such as controlled ones---might contribute to clarify the effort saved with RAID. The inspection time of the float window is also a threat to validity, since a developer can have left the window opened beyond the time used in the review.

\section{Related Work}
\label{sec:related-work}

Due to its importance, the literature presents several studies exploring refactoring-related problems. For instance, there are studies on motivations to perform refactoring operations~\cite{fse2016danilo, Mazinanian:2017:OOPSLA, tsantalis2013,Pantiuchina:2020}, challenges and benefits~\cite{Kim:2012:FSE, Kim:2014:TSE}, the impact on software quality~\cite{Bibiano:esem2019:BatchRefactoring, Chaparro:ICSME:2014, lacerda2020code, Catolino:2020:ICSE}, security~\cite{abid2020does, mumtaz2018empirical, maruyama2008}, and also on tools to assist developers during refactoring tasks~\cite{ge2011benefactor, Shen:OOPSLA:2019, Alves:2014:FSE, jss-2018-jmove, alizadeh2019refbot, Lahiri2012, saner2019-gocity}.
Besides, various studies focused on the identification of refactoring operations through analysis of version control systems~\cite{tsantalis2018, tsantalis2020,silva2017, silva2020,brito2020,Shen:OOPSLA:2019}. 

Recently, Tsantalis {\em et al.}~\cite{tsantalis2018, tsantalis2020} presented RefactoringMiner that identifies 40 distinct refactoring types in Java. 
The authors evaluate the last tool version with 7,226 refactoring instances, reporting  a high precision (99.6\%) and recall (94\%)~\cite{tsantalis2020}. 
Currently, RAID does not use this tool. 
However, in the future we plan to extend RAID to also work with RefactoringMiner.


Silva {\em et al.}~\cite{silva2017, silva2020} proposed RefDiff, a multi-language tool to detect refactoring activities. 
The last version identifies 13 distinct refactoring types, and it supports three programming languages (i.e., Java, JavaScript, and C). 
The authors also evaluated the tool, reporting a right accuracy for these three languages. Specifically, the authors reported 96\% of precision and 80\% of recall for Java;  91\% of precision and 88\% of recall for JavaScript; and 88\% of precision and 91\% of recall for C.
 RefDiff also supports other programming languages via a plugin system.
 Recently, we proposed the RefDiff for Go plugin.
 We evaluated our plugin with six well-known open-source projects, reporting a precision of 92\% and 80\% of recall~\cite{brito2020}.
 RAID---as proposed and evaluated in this paper---relies on this plugin extension for Go.

Textual diffs also present issues when merging changes in version control systems. Bo {\em et al.}~\cite{Shen:OOPSLA:2019} proposed a graph-based refactoring-aware algorithm to detect and resolve refactoring-related conflicts. The algorithm was evaluated with 1,070 merge scenarios from popular open source Java projects and archived a 58.90\% reduction of merge conflicts comparing with GitMerge and 11.84\% over jFSTMerge.

The benefits of refactoring-aware code reviews have also been explored in the literature. For example, Ge {\em et al.}~\cite{ge2017} presented a formative study with 35 developers to investigate the motivation and challenges of reviewing refactorings during code review. They report that 94\% (33) of the study participants consider that refactorings can slow down the review of non-refactoring changes. Furthermore, the automatic identification of refactorings may assist code reviewers for 91\% of the participants (32).  The authors also presented a refactoring-aware tool called ReviewFactor, which provides a separation of refactorings and non-refactorings changes for the purpose of code review. This tool separates code reviews in two steps: first, non-refactored code should be reviewed using a specific interface that is part of an IDE; after that, reviewers should focus on the refactored code. With RAID we decided to follow a different approach by seamlessly integrating our tool to a state-of-the-practice code review workflow. For this purpose, RAID provides the ``R'' buttons and also the floating windows with detailed data about refactorings. It is also relevant to mention that  refactorings are usually followed by minor edits in the changed code~\cite{fse2016danilo,silva2020,tsantalis2020}, as  illustrated in Figures \ref{fig:move-raid} and \ref{fig:extract-raid}. However, it is not clear how ReviewFactor handles such interleaved operations, which might for example require a duplicated effort by code reviewers. 

There are also open source projects proposing refactoring-aware tools. The tool proposed by Tsantalis {\em et al.}~\cite{tsantalis2020}, called Refactoring Aware Commit Review, identifies refactorings in open source Java projects and lists the refactorings activities on GitHub diffs. RefactorInsight is another tool for visualizing refactorings through the history of commits, but within a IDE for Java and Kotlin.\footnote{\url{https://github.com/JetBrains-Research/RefactorInsight}} Both tools instrument textual diff results with a list of refactoring operations, which are detected by RefactoringMiner~\cite{tsantalis2020}. In Refactoring Aware Commit Review, after reviewers click on a given list element that describes a refactoring, they are directed to its specific line in the right side of GitHub's diff. Therefore, both tools do not instrument diff lines with refactoring data, unlike RAID that seamlessly provides the ``R'' buttons, which are added to both sides of a diff (see Figure~\ref{fig:r-annotation}). Also, these tools do not provide detailed information about Move Function refactorings---with the moved code appearing side by side, as in Figure~\ref{fig:move-raid}---or detailed information about Extract Function refactorings---by highlighting the extracted code and the textual differences in the source method, as illustrated in Figure \ref{fig:extract-raid}. 
Finally, unlike RAID, both tools are not integrated with a CI server, which is critical to provide refactoring data right after pull requests are submitted.

Other studies on refactoring-aware code reviews are summarized in a survey by Coelho {\em et al.}~\cite{Coelho:IWoR:2019}.

\section{Conclusion}
\label{sec:conclusion}

In this paper, we presented RAID, a refactoring-aware tool that instruments GitHub diff with refactoring information. We also conducted a field experiment with eight professional developers during three months and we concluded that RAID can reduce the cognitive effort required  for reviewing refactorings when using textual diffs. In addition, our study reports a reduction in the number of lines required for reviewing such operations. In the case of move refactorings, the number of lines decrease form 14.5 to 2 lines (median values); and from 113 to 55 lines in the case of extractions. 

As future work, we plan to configure and evaluate RAID with other refactoring detection tools, such as RefactoringMiner~\cite{tsantalis2020}. We also plan to provide plugin support for other browsers, such as Mozilla Firefox and Safari. Finally, we intend to evaluate our approach using a controlled experiment to better understand the benefits and impacts of the tool.

The tool is public available at GitHub.\footnote{\url{https://github.com/rodrigo-brito/refactoring-aware-diff}}

\section*{Acknowledgments}

\noindent We thank the eight developers who participated in our study and shared their ideas and practices about refactoring and code review. This research is supported by grants from CAPES, FAPEMIG, and CNPq.


\bibliographystyle{IEEEtran}
\bibliography{document}

\end{document}